\documentclass[sigconf]{acmart}
\usepackage{url}
\usepackage[tight,footnotesize]{subfigure}
\usepackage{balance}
\usepackage{multirow}
\usepackage{fancybox}
\usepackage{mathrsfs}
\usepackage{wrapfig}
\usepackage{graphicx}
\usepackage{amsmath}
\usepackage{mathtools}
\usepackage[mathscr]{euscript}
\usepackage{algorithm}
\usepackage{algorithmic}

\usepackage{amsthm}

\theoremstyle{definition}
\theoremstyle{remark}

\graphicspath{{figures/}}
\usepackage{ctable}
\usepackage{bbm}
\usepackage{verbatim}
\usepackage{comment}

\usepackage{booktabs} 

\usepackage{fancyhdr,lipsum}
\copyrightyear{2017}
\acmYear{2017}
\setcopyright{acmlicensed}
\acmConference{CIKM'17 }{November 6--10, 2017}{Singapore, Singapore}\acmPrice{15.00}\acmDOI{10.1145/3132847.3132873}
\acmISBN{978-1-4503-4918-5/17/11}

\begin{document}

\thispagestyle{plain}

\title{Name Disambiguation in Anonymized Graphs using Network Embedding\footnote{This research is sponsored by Mohammad Al Hasan's NSF CAREER Award (IIS-1149851) and also, by a research grant from CareerBuilder.}}

\author{Baichuan Zhang}
\affiliation{%
  \institution{Purdue University}
  \city{West Lafayette} 
  \state{IN, USA} 
}
\email{zhan1910@purdue.edu}

\author{Mohammad Al Hasan}
\orcid{orcid.org/0000-0002-8279-1023}
\affiliation{%
  \institution{Indiana University Purdue University Indianapolis}
  \city{Indianapolis} 
  \state{IN, USA} 
}
\email{alhasan@cs.iupui.edu}

\begin{abstract}

In real-world, our DNA is unique but many people share names. This phenomenon
often causes erroneous aggregation of documents of multiple persons who are
namesake of one another. Such mistakes deteriorate the performance of document
retrieval, web search, and more seriously, cause improper attribution of credit
or blame in digital forensic. To resolve this issue, the name disambiguation
task is designed which aims to partition the documents associated with a name
reference such that each partition contains documents pertaining to a unique
real-life person. Existing solutions to this task substantially rely on feature
engineering, such as biographical feature extraction, or construction of
auxiliary features from Wikipedia. However, for many scenarios, such features
may be costly to obtain or unavailable due to the risk of privacy violation. In
this work, we propose a novel name disambiguation method. Our proposed method
is non-intrusive of privacy because instead of using attributes pertaining to a
real-life person, our method leverages only relational data in the form of
anonymized graphs.  In the methodological aspect, the proposed method uses a
novel representation learning model to embed each document in a low
dimensional vector space where name disambiguation can be solved by a
hierarchical agglomerative clustering algorithm. Our experimental results
demonstrate that the proposed method is significantly better than the existing
name disambiguation methods working in a similar setting.  

\end{abstract}

%
%



\maketitle

%
%

\section{Introduction}

%
%
Name
disambiguation~\cite{Han.Giles.ea:04,Cen.Luo.ea:13,Zhang.Tang.ea:07,Zhang.Saha.ea:14,Zhang.Dundar.ea:16}
is an important problem, which has numerous applications in information
retrieval, counter-terrorism, and bibliographic data analysis. In information
retrieval, name disambiguation is critical for sanitizing search results of
ambiguous queries. For example, an online search query for ``Michael Jordan"
may retrieve pages of former US basketball player, the pages of UC Berkeley
machine learning professor, and the pages of other persons having that name,
and name disambiguation is necessary to split those pages into homogeneous
groups.  In counter-terrorism, such an exercise is essential before inserting a
person's profile in a law enforcement database; failing to do so may cause
severe trouble to many innocent persons who are namesakes of a potential criminal.
Evidently, name disambiguation is particularly important in the fields of
bibliometrics and library science. This is due to the fact that many distinct
authors share the same name reference as the first name of an author is
typically written in abbreviated form in the citation of many scientific
articles. Thus, bibliographic servers that maintain such data may mistakenly
aggregate the articles from multiple scholars (sharing the same name) into a
unique profile in some digital repositories. For an example, the Google scholar profile
associated with the name ``Yang Chen" (GS)~\footnote{\url{https://scholar.google.com/citations?user=gl26ACAAAAAJ&hl=en}}
is verified as the profile page of a Computer Graphics PhD candidate at Purdue University, but based 
on our labeling, more than
20 distinct persons' publications are mixed under that profile mistakenly. 
Such mistakes in library science
over- or under-estimate a researcher's citation related impact metrics. 

Due to its importance, the name disambiguation task has attracted substantial
attention from information retrieval and data mining communities. However, the
majority of existing
solutions~\cite{Bunescu.Pasca:06,Cen.Luo.ea:13,Han.Sun.ea:11,Hoffart.Yosef.ea:11}
for this task use biographical features such as name, address, institutional
affiliation, email address,  and homepage. Also, contextual features such as
collaborator, community affiliation, and external data source such as Wikipedia
are used in some works~\cite{Hoffart.Yosef.ea:11,Han.Zhao:09}. Using
biographical features is acceptable for disambiguation of authors in
bibliometrics domain, but in many scenarios, for example in the national
security related applications, biographical features are hard to obtain, or
they may even be illegal to obtain unless a security analyst has the
appropriate level of security clearance. Besides, in real-world social networks
(e.g., Twitter, Facebook, and LinkedIn), some users may choose a strict privacy
setting that restricts the visibility of their profile information and posts.
For such privacy-preserving scenarios, many existing name disambiguation
techniques~\cite{Han.Sun.ea:11,Hoffart.Yosef.ea:11,Tang.Fong.ea:12,Wang.Tang.ea:11,Han.Giles.ea:04},
which compute document similarity using biographical attributes are not
applicable.

In recent years, a few works have emerged where name disambiguation task in
privacy-preserving setting has been
considered~\cite{Hermansson.Kerola.ea:13,Zhang.Saha.ea:14}. These works use
relational data in the form of an anonymized person-person collaboration
graph, and solve name disambiguation by using graph topological features.  Thus
they preserve the privacy of a user. Authors of~\cite{Hermansson.Kerola.ea:13}
use graphlet kernels based classification model and the authors
of~\cite{Zhang.Saha.ea:14} use Markov clustering based unsupervised
approach.  However, both of these works only consider a binary classification
task, predicting whether a given person-node in the graph is ambiguous or
non-ambiguous.  This is far from a traditional name disambiguation task which
partitions the records pertaining to a given name reference into different
groups, each belonging to a unique person. Another limitation of the existing
works is that they only utilize the person-person collaboration network, which
does not generally yield a good disambiguation performance. There are other
information, such as person-document association information and
document-document similarity information, which can also be exploited for
obtaining improved name disambiguation, yet preserving the user's privacy. 

In this work, we solve the name disambiguation task by using only relational
information. For a given name reference, our proposed method pre-processes the
input data as three graphs: person-person graph representing collaboration
between a pair of persons, person-document graph representing association of a
person with a document and document-document similarity graph. These graphs are
appropriately anonymized, as such, the vertices of these graphs are represented
by a unique pseudo-random identifier. Nodal features (such as, biographical
information of a person-node, or keywords of a document-node)  of any of the
above three graphs are not used, which makes the proposed method
privacy-preserving. 

In the graph representation, the name disambiguation task becomes a graph
clustering task of the document-document graph, with the objective that each
cluster contains documents pertained to a unique real-life person. A traditional method to 
cluster a homogeneous network cannot
facilitate information exchange among the three graphs, so we propose a novel
representation learning model, which embeds the  vertices of these graphs into
a shared low dimensional latent space by using a joint objective function.  The
objective function of our representation learning task utilizes pairwise
similarity ranking which is different from the typical
objective functions used in the existing 
document embedding methods, such as LINE~\cite{Tang.Wang.ea:15} and PTE~\cite{Tang.Qu.ea:15}; the latter ones are
based on  K-L divergence between empirical similarity distribution and embedding similarity distribution. K-L divergence works over the entire
distribution vector and it works well for document labeling or
topic modeling, but not so for clustering. On the other hand, our objective function is better suited
for a downstream clustering task because it directly optimizes the pairwise
distance between similar and dissimilar documents,
thus making the document vectors disambiguation-aware
in the embedded space,
as such, a traditional hierarchical clustering of the
vectors in the embedded space generates excellent name disambiguation
performance. Experimental comparison with several state-of-the-art
name disambiguation methods---both traditional and network 
embedding-based---show that the 
proposed method is significantly better than the existing
methods on multiple real-life name disambiguation 
datasets.\\

The key contributions of this work are summarized as below:

\begin{enumerate}

\item We solve the name disambiguation task by using only linked data from
network topological information.  The work is motivated by the growing demand
for big data analysis without violating the user privacy in security sensitive domains.

\item We propose a network embedding based solution that leverages linked
structures of a variety of anonymized networks in order to represent each
document into a low-dimensional vector space for solving the name
disambiguation task. To the best of our knowledge, our work is the first one to
adopt a representation learning framework for name disambiguation in anonymized
graphs.

\item For representation learning, we present a novel pairwise ranking based objective, which is particularly suitable for solving the name disambiguation task
by clustering.

\item We use two real-life bibliographic datasets for evaluating the
disambiguation performance of our solution. The results  demonstrate the
superiority of our proposed method over the state-of-the-art methodologies for name disambiguation in privacy-preserving setup. 

\end{enumerate}

\section{Related Work}

There exist a large number of works on name disambiguation~\cite{Han.Giles.ea:04,Cen.Luo.ea:13}. In terms of methodologies,  
existing works have considered
supervised~\cite{Bunescu.Pasca:06, Han.Giles.ea:04}, unsupervised~\cite{Han.Zha.ea:05,Cen.Luo.ea:13}, and probabilistic relational models~\cite{Tang.Fong.ea:12,Song.Huang.ea:07,Zhang.Tang.ea:07}.
In the supervised setting, Han et al.~\cite{Han.Giles.ea:04} proposed supervised name disambiguation methodologies by utilizing Naive Bayes and 
SVM. In these works, a distinct real-life
entity can be considered as a class, and the objective is to classify each record to one of the classes.
For the unsupervised name disambiguation, the records are partitioned into several clusters with the goal of
obtaining a partition where each cluster contains records from a unique entity. For example, Han et al.\cite{Han.Zha.ea:05} used $K$-way spectral clustering for name disambiguation in bibliographical data. 
Recently, probabilistic relational models, especially graphical models have also been
considered for the name disambiguation task. For instance,~\cite{Tang.Fong.ea:12} proposed to use Markov
Random Fields to address name disambiguation in a unified probabilistic framework. 

Most existing solutions to the name disambiguation task use either biographical attributes, or auxiliary features that are collected 
from external sources. However, the attempt of extracting biographical or external data sustains the risk of privacy violation. 
To address this issue, a few works~\cite{Malin:05,Zhang.Saha.ea:14,Hermansson.Kerola.ea:13,Zhang.Hasan.ea:15} have considered name disambiguation using anonymized graphs 
without leveraging the node attributes. 
The central idea of this type of works is to exploit graph topological features to solve the name disambiguation problem without intruding user privacy through the collection of bibliographical attributes.
For example, authors in~\cite{Hermansson.Kerola.ea:13} characterized the similarity between two nodes based on their local neighborhood structures using graph kernels and solved the name disambiguation problem using SVM. However, the major drawback of the proposed method in~\cite{Hermansson.Kerola.ea:13}
is that it  can only detect entities that should be disambiguated, but fails to further partition the documents into their corresponding homogeneous groups. Authors in~\cite{Zhang.Saha.ea:14,Zhang.Hasan.ea:15} proposed an unsupervised solution to name disambiguation in an anonymized graph by exploiting the time-stamped network topology around a vertex. However, it also suffers from the similar issue as described above. 

%
%
Our proposed solution utilizes a network representation learning based
approach~\cite{Chang.Han.ea:15,Perozzi.Skiena.ea:14,Tang.Qu.ea:15,Tang.Wang.ea:15,
Grover.Leskovec:16,Cao.Lu.ea:15, Chen.Sun:17,Wang.Liu.CIKM, acl-17}--- a rather recent
development in machine learning. Many of these methods are inspired
by word embedding based language model~\cite{Mikolov.Chen.ea:13}.  Different
from traditional graph embedding methods, such as  
Laplacian Eigenmaps~\cite{Chen.Hero.ea:16,chen2016incremental},
the recently proposed network embedding methods, such as
DeepWalk~\cite{Perozzi.Skiena.ea:14}, LINE~\cite{Tang.Wang.ea:15},
PTE~\cite{Tang.Qu.ea:15}, and Node2Vec~\cite{Grover.Leskovec:16}, are more
scalable and have shown better performance in node classification and link
prediction tasks. Among these works, LINE~\cite{Tang.Wang.ea:15} finds embedding
of documents by using document-document similarity matrix, whereas our work
uses multiple networks and performs a joint learning. PTE~\cite{Tang.Qu.ea:15} 
performs a joint learning of multiple input graphs, but PTE needs labeled data.
Finally, the embedding formulation and optimization of our proposed method is
different than LINE or PTE. Specifically, we use a ranking based loss function
as our objective function whereas mostly all the existing methods use K-L divergence
based objective function.

\section{Problem Formulation}\label{sec:pf}

We first introduce notations used in this paper. Throughout the paper, bold uppercase letter (e.g., $\mathbf{X}$) denotes a matrix, 
bold lowercase letter such as $\mathbf{x}_{i}$ denotes a column vector, and $(\cdot)^{T}$ denotes vector transpose. $\lVert \mathbf{X} \rVert_{F}$ is the Frobenius norm of matrix $\mathbf{X}$.  Calligraphic uppercase letter (e.g., $\mathcal{X}$) is used to denote a set and 
$|\mathcal{X}|$ is the cardinality of the set $\mathcal{X}$.

For a given name reference $a$, we denote $\mathcal{D}^{a} = \{d_{1}^{a},
d_{2}^{a}, ..., d_{N}^{a}\}$ to be a set of $N$ documents with which $a$ is
associated and $\mathcal{A}^{a} = \{a_{1}, a_{2}, ..., a_{M} \}$ is the
collaborator set of $a$ in $\mathcal{D}^{a}$, where $a \not \in \mathcal{A}^a$.
If there is no ambiguity we remove the superscript $a$ in the notations of both
$\mathcal{D}^{a}$ and $\mathcal{A}^{a}$ and refer the terms as $\mathcal{D}$
and $\mathcal{A}$, respectively.  For illustration, in bibliographic field,
$\mathcal{D}$ can be the set of scholarly publications where $a$ is one of the
authors and $A$ is the set of $a$'s coauthors.   In real-life, the given name
reference $a$ can be associated with multiple persons (say $L$) all sharing the
same name. The task of name disambiguation is to partition $\mathcal{D}$ into
$L$ disjoint sets such that each partition contains documents of a unique
person entity with name reference $a$.

\begin{figure}
\centering
\includegraphics[width=70mm]{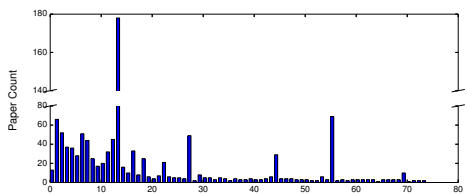}
\caption{Paper Count Distribution of ``S Lee''}
\label{fig:dist-slee}
\vspace{-0.1in}
\end{figure}

Though it may appear as a simple clustering problem, name disambiguation is challenging on
real-life data. This is due to the fact that it requires solving a highly class-imbalanced clustering task,
as the number of documents associated with a distinct person follows a power-law distribution. 
We demonstrate it through an example from the bibliographic domain. In Figure~\ref{fig:dist-slee},
we show a histogram of paper counts of various real-life persons named ``S Lee'' in CiteSeerX~\footnote{\url{http://citeseerx.ist.psu.edu/index;jsessionid=4A26742FADC605600567F493C2D7825E}}.
As we can observe, there are a few real-life authors (dominant entities) with the name  ``S Lee'' 
to whom the majority of the publications belong. Only a few publications belong to each of the remaining real-life authors with name ``S Lee''. Due to this severe class imbalance issue, majority of traditional clustering methods
perform poorly on this task. Sophisticated machine learning models, like the one we propose below are needed 
for solving this task. This example is from bibliographic domain, but power-law distribution of possession 
is common in every aspect of real-life, so we expect this challenge to hold in 
other domains as well.

In this study, we investigate the name disambiguation problem in a restricted
setup, where bibliographical features and information from external sources are
not considered so that  the risk of privacy violation can be alleviated.
Instead, we formulate the problem using graphs in which each node has been
assigned an anonymized identifier, and network topological structure is the
only information available. Specifically, our solution encodes the local
neighborhood structures accumulated from three different networks into a
proposed network embedding model, which generates a $k$-dimensional vector
representation for each document. The networks are person-person network,
person-document network, and linked document network, which we formally define
as below: 

\begin{definition}[Person-Person Network] {\em For a given name reference $x$,
the person-person network, denoted as $G_{pp} = (\mathcal{A}^x, E_{pp})$,
captures collaboration between a pair of persons within the collection of documents
associated with $x$. $\mathcal{A}^x$ is the collaborator set, and $e_{ij} \in E_{pp}$
represents the edge between the persons, $a_{i}$ and $a_{j}$, who
collaborated in at least one document. The weight $w_{ij}$ of the edge $e_{ij}$
is defined as the number of distinct documents in which $a_{i}$ and $a_{j}$ have collaborated.}
\label{def:1} \end{definition}


The person-person network is important because the inter-person acquaintances
represented by collaboration relation can be used to discriminate the set of
documents of multiple real-life persons. However, the collaboration network
does not account for the fact that the documents associated with the same real-life
person are inherently similar; person-document network and document-document
network cover for this shortcoming. 

\begin{definition}[Person-Document Network]{\em Person-Document Network,
represented as $G_{pd} = (\mathcal{A \cup D}, E_{pd})$, is a bipartite network
where $\mathcal{D}$ is the set of documents with which the name reference $a$
is associated and $\mathcal{A}$ is the set of collaborators of $a$ over all the
documents in $\mathcal{D}$. $E_{pd}$ is the set of edges between persons and
documents. The edge weight $w_{ij}$ between a person node $a_{i}$ and document
$d_{j}$ is simply defined as the number of times $a_{i}$ appears in document
$d_{j}$.  For a bibliographic dataset, $a_i$ is simply an author of the
document $d_j$ and the weight $w_{ij} = 1$.  } \label{def:2} \end{definition}

\begin{definition}[Linked Document Network]{\em Document-Document Network,
represented as $G_{dd} = (\mathcal{D}, E_{dd})$,  where each vertex $d_{i} \in
\mathcal{D}$ is a document. If two documents $d_{i}$ and $d_{j}$ are similar
(more discussion is forthcoming), we build an edge between them represented as
$e_{ij} \in E_{dd}$.  } \label{def:3} \end{definition}

There are several ways document-document similarity 
can be captured. For instance, one can find word co-occurrence between different 
documents to compute this similarity. However, we refrained from using word co-occurrence 
due to the privacy concern as sometimes a list of
a set of unique words can reveal the identity of a person~\cite{Zhang.Noman.ea:17}. 
Instead we define document-document similarity 
through a combination of person-person and person-document relationships. 
Two documents are similar if the intersection
of their collaborator-sets is large (by using person-document relationship) or if 
the intersection of one-hop neighbors of their collaborator-sets is large (by using both 
person-document and person-person relationships). 

The above definition of document similarity captures two important patterns
which facilitate effective name disambiguation by document clustering. First,
there is a high chance for two documents to be authored by the same real-life
person, if they have a large number of overlapping collaborators. Second, even if
they do not have any overlapping collaborators, large overlap in the neighbors of
their collaborators signals that the documents are most likely authored by the
same person. For both cases, these two documents should be placed in close
proximity in the embedded space.  Mathematically, we denote
$\mathcal{A}_{d_{i}}^{1}$ as the collaborator set of $d_{i}$. Furthermore,
$\mathcal{A}_{d_{i}}^{2}$ is the set of collaborators by extending
$\mathcal{A}_{d_{i}}^{1}$ with all neighbors of the persons in
$\mathcal{A}_{d_{i}}^{1}$, namely $\mathcal{A}_{d_{i}}^{2} =
\mathcal{A}_{d_{i}}^{1} \cup \{\mathcal{NB}_{G_{pp}}(b) \}_{b \in
\mathcal{A}_{d_{i}}^{1}}$, where $\mathcal{NB}_{G_{pp}}(b)$ is the set of
neighbors of node $b$ in person-person network $G_{pp}$. Then the document
similarity between $d_{i}$ and $d_{j}$ in the graph $G_{dd}$ is simply defined as $w_{ij} =
|\mathcal{A}_{d_{i}}^{2} \cap \mathcal{A}_{d_{j}}^{2}|$.

Based on our problem formulation, the name disambiguation solution
consists of two phases: (1) document representation  (2) disambiguation.
We discuss them as below:

Given a name reference $a$, its associated document set 
$\mathcal{D}^{a}$ (which we want to disambiguate) and
the collaborator set $\mathcal{A}^{a}$, the document representation phase 
first constructs corresponding person-person network $G_{pp}$, 
person-document bipartite network $G_{pd}$, and linked document 
network $G_{dd}$. Then our proposed document representation model 
combines structural information from these three networks to 
generate a $k$-dimensional document embedding matrix 
$\textbf{D} = [\mathbf{d}_{1}^{T}, ..., \mathbf{d}_{N}^{T}] \in {\rm I \!R}^{N \times k}$. 

Then the disambiguation phase takes the document embedding matrix $\textbf{D}$
as input and applies the hierarchical agglomerative clustering (HAC) with group
average merging criteria to partition $N$ documents in $\mathcal{D}^{a}$ into
$L$ disjoint sets with the expectation that each set is composed of documents
of a unique person entity sharing the name reference $a$. At this stage, $L$ is
a user-defined parameter which we match with the ground truth during the
evaluation phase. In real-life though, a user needs to tune the parameter $L$
which can easily be done with HAC, because HAC provides hierarchical
organization of clusters at all levels starting from a single cluster upto the
case of single-instance cluster, and a user can recover clustering for any
value of $L$ as needed without additional cost. Also, across different $L$
values the cluster assignment of HAC is consistent (i.e., two instances that are in
the same cluster for some $L$ value will remain in the same cluster for any
smaller $L$ value), which further helps in choosing an appropriate $L$ value.

\section{Method}\label{sec:method}

In this section, we discuss our proposed representation learning model for name
disambiguation. Our goal is to encode the local neighborhood structures
captured by the three networks (see Definitions~\ref{def:1} ~\ref{def:2}
~\ref{def:3}) into the $k$-dimensional document embedding matrix with strong
name disambiguation ability.

\subsection{Model Formulation}

The main intuition of our network embedding model is that neighboring nodes in
a graph should have more similar vector representation in the embedding
space than non-neighboring nodes. For instance, in linked document network,
the affinity between two neighboring vertices $d_{i}$ and $d_{j}$, i.e.,
$e_{ij} \in G_{dd}$ should be larger than the affinity between two
non-neighboring vertices $d_{i}$ and $d_{t}$, i.e., $e_{it} \not \in G_{dd}$.
The affinity score between two nodes $d_{i}$ and $d_{j}$ in $G_{dd}$ can be
calculated as the inner product of their corresponding embedding
representations, denoted as $S_{ij}^{dd} = \mathbf{d}_{i}^{T}\mathbf{d}_{j}$.
More specifically, we model the probability of preserving ranking order
$S_{ij}^{dd} > S_{it}^{dd}$ using the logistic function $\sigma(x) = \frac{1}{1
+ e^{-x}}$. Mathematically,

\begin{equation}
P\big(S_{ij}^{dd} > S_{it}^{dd} | \mathbf{d}_{i}, \mathbf{d}_{j}, \mathbf{d}_{t}\big) = \sigma\big(S_{ijt}^{dd}\big)
\label{eq:1}
\end{equation}
where $S_{ijt}^{dd}$ is defined as below:

\begin{eqnarray}
S_{ijt}^{dd} &=& S_{ij}^{dd} - S_{it}^{dd} \nonumber \\
&=& \mathbf{d}_{i}^{T}\mathbf{d}_{j} - \mathbf{d}_{i}^{T} \mathbf{d}_{t} \nonumber \\
\end{eqnarray}

As we observe from Equation~\ref{eq:1}, the larger $S_{ijt}^{dd}$, the more likely ranking order $S_{ij}^{dd} > S_{it}^{dd}$ is preserved. By assuming all the ranking orders generated from the linked document network $G_{dd}$ to be independent, the probability $P(> | \mathbf{D})$ of all the ranking orders being preserved given the document embedding matrix $\mathbf{D} \in {\rm I \!R}^{N \times k}$ is defined as below:

\begin{eqnarray}
P(> | \mathbf{D}) &=& \prod_{\substack{(d_{i}, d_{j}) \in \mathcal{P}_{G_{dd}} \\ (d_{i}, d_{t}) \in \mathcal{N}_{G_{dd}}}} P\big(S_{ij}^{dd} > S_{it}^{dd} | \mathbf{d}_{i}, \mathbf{d}_{j}, \mathbf{d}_{t}\big) \nonumber \\
&=& \prod_{\substack{(d_{i}, d_{j}) \in \mathcal{P}_{G_{dd}} \\ (d_{i}, d_{t}) \in \mathcal{N}_{G_{dd}}}}\sigma\big(S_{ijt}^{dd}\big) \nonumber \\
&=& \prod_{\substack{(d_{i}, d_{j}) \in \mathcal{P}_{G_{dd}} \\ (d_{i}, d_{t}) \in \mathcal{N}_{G_{dd}}}}\sigma\big( S_{ij}^{dd} - S_{it}^{dd}\big) \nonumber \\
\label{eq:2}
\end{eqnarray}
where $\mathcal{P}_{G_{dd}}$ and $\mathcal{N}_{G_{dd}}$ are positive and negative training sets in $G_{dd}$.

From the Equation~\ref{eq:2}, the goal is to seek the document latent representation $\mathbf{D}$ for all nodes in linked document network $G_{dd}$, which maximizes $P(> | \mathbf{D})$. For the computational convenience, we minimize the following sum of negative log-likelihood objective, which is shown as follows:

\begin{eqnarray}
OBJ_{dd} &=& \underset{\mathbf{D}}{\text{min}} -\ln P(>|\mathbf{D}) \nonumber \\
&=& -\sum_{\substack{(d_{i}, d_{j}) \in \mathcal{P}_{G_{dd}} \\ (d_{i}, d_{t}) \in \mathcal{N}_{G_{dd}}}} \ln P\big(S_{ij}^{dd} > S_{it}^{dd} | \mathbf{d}_{i}, \mathbf{d}_{j}, \mathbf{d}_{t}\big) \nonumber \\
&=& -\sum_{\substack{(d_{i}, d_{j}) \in \mathcal{P}_{G_{dd}} \\ (d_{i}, d_{t}) \in \mathcal{N}_{G_{dd}}}} \ln \sigma(S_{ijt}^{dd}) \nonumber \\
&=& -\sum_{\substack{(d_{i}, d_{j}) \in \mathcal{P}_{G_{dd}} \\ (d_{i}, d_{t}) \in \mathcal{N}_{G_{dd}}}} \ln \sigma\big(S_{ij}^{dd} - S_{it}^{dd}\big) \nonumber \\
\label{eq:3}
\end{eqnarray}

The formulation shown in Equation~\ref{eq:3} constructs a probabilistic
framework for distinguishing between neighbor nodes and non-neighbor nodes in a linked
document network by preserving a ranking order objective function. 

Using the identical argument, the objective functions for capturing person-person and person-document relations are given as below:

\begin{eqnarray}
OBJ_{pp} &=& \underset{\mathbf{A}}{\text{min}} -\ln P(>|\mathbf{A}) \nonumber \\
&=& -\sum_{\substack{(a_{i}, a_{j}) \in \mathcal{P}_{G_{pp}} \\ (a_{i}, a_{t}) \in \mathcal{N}_{G_{pp}}}} \ln \sigma(S_{ij}^{pp} - S_{it}^{pp}) \nonumber \\
\label{eq:4}
\end{eqnarray}
\begin{eqnarray}
OBJ_{pd} &=& \underset{\mathbf{A}, \mathbf{D}}{\text{min}} -\ln P(>|\mathbf{A}, \mathbf{D}) \nonumber \\
&=& -\sum_{\substack{(d_{i}, a_{j}) \in \mathcal{P}_{G_{pd}} \\ (d_{i}, a_{t}) \in \mathcal{N}_{G_{pd}}}} \ln \sigma(S_{ij}^{pd} - S_{it}^{pd}) \nonumber \\
\label{eq:5}
\end{eqnarray}

where $\mathbf{A} \in {\rm I \!R}^{M \times k}$ can be thought as the person
embedding matrix and $M$ is the number of persons in the collaborator set
$\mathcal{A}$. $S_{ij}^{pp}$ represents the affinity score between two nodes
$a_{i}$ and $a_{j}$ in collaboration graph $G_{pp}$, and $S_{ij}^{pd}$ denotes
the affinity score between two nodes $d_{i}$ and $a_{j}$ in heterogeneous
bipartite graph $G_{pd}$. Finally, $\mathcal{P}_{G_{pp}}$ and
$\mathcal{N}_{G_{pp}}$ are positive and negative training sets in $G_{pp}$,
$\mathcal{P}_{G_{pd}}$ and $\mathcal{N}_{G_{pd}}$ are positive and negative
training sets in $G_{pd}$ respectively. 

The goal of proposed network embedding framework is to unify these three types
of relations together, where the person and document vertices are shared across
these three networks. An intuitive manner is to collectively embed these three
networks, which can be achieved by minimizing the following objective function:

\begin{equation}
OBJ = \underset{\mathbf{A}, \mathbf{D}}{\text{min}}  -OBJ_{pp} - OBJ_{pd} - OBJ_{dd} + \lambda Reg(\mathbf{A}, \mathbf{D}) 
\label{eq:6}
\end{equation}
where $\lambda Reg(\mathbf{A}, \mathbf{D}$) in Equation~\ref{eq:6} is a $l_{2}$-norm regularization term to prevent the model from overfitting. Here for the computational convenience, we set $Reg(\mathbf{A}, \mathbf{D})$ as $\lVert \mathbf{A} \rVert_{F}^{2} + \lVert \mathbf{D} \rVert_{F}^{2}$. Such pairwise ranking loss objective is in the similar spirit to the Bayesian Personalized Ranking~\cite{zhang2016trust,choudhury2017nous}, which aims
to predict the interaction between users and items in recommender system domain. 

%
%

\subsection{Model Optimization~\label{sec:modellearning}}

We use stochastic gradient descent (SGD) algorithm for
optimizing Equation~\ref{eq:6}. Specifically, in each step we sample the
training instances involved in person-person, person-document, and
document-document relations accordingly. The sampling strategy of positive
instances is based on edge sampling \cite{Tang.Qu.ea:15}. Specifically, for
example, in linked document network $G_{dd}$, given an arbitrary node $d_{i}$,
we sample one of its neighbors $d_{j}$, i.e., $(d_{i}, d_{j}) \in
\mathcal{P}_{G_{dd}}$, with the probability proportional to the edge weight for
the model update. On the other hand, for sampling of negative instances, we
utilize uniform sampling technique. 
In particular, given the sampled node
$d_{i}$, we sample an arbitrary negative instance $d_{t}$ uniformly, namely
$(d_{i}, d_{t}) \in \mathcal{N}_{G_{dd}}$.

Therefore given a sampled triplet $(d_{i}, d_{j}, d_{t})$ with $(d_{i}, d_{j}) \in \mathcal{P}_{G_{dd}}$ and $(d_{i}, d_{t}) \in \mathcal{N}_{G_{dd}}$, using the chain rule and back-propagation, the gradient of the objective function $OBJ$ in Equation~\ref{eq:6} w.r.t. $\mathbf{d}_{i}$ can be computed as below:

\begin{eqnarray}
\frac{\partial OBJ}{\partial \mathbf{d}_{i}} &=& -\frac{\partial \ln\sigma\big(S_{ij}^{dd} - S_{it}^{dd}\big)}{\partial \mathbf{d}_{i}} + 2\lambda \mathbf{d}_{i} \nonumber \\ 
&=& -\frac{\partial \ln\sigma\big(S_{ij}^{dd} - S_{it}^{dd}\big)}{\partial \sigma\big(S_{ij}^{dd} - S_{it}^{dd}\big)} \times \frac{\partial\sigma\big(S_{ij}^{dd} - S_{it}^{dd}\big)}{\partial\big(S_{ij}^{dd} - S_{it}^{dd}\big)} \nonumber\\
&& \times \frac{\partial\big(S_{ij}^{dd} - S_{it}^{dd}\big)}{\partial \mathbf{d}_{i}} + 2\lambda \mathbf{d}_{i} \nonumber \\
&=& - \frac{1}{\sigma\big(S_{ij}^{dd} - S_{it}^{dd}\big)} \times \sigma\big(S_{ij}^{dd} - S_{it}^{dd}\big) \nonumber\\
&& \Big(1-\sigma\big(S_{ij}^{dd} - S_{it}^{dd}\big)\Big)\times(\mathbf{d}_{j} - \mathbf{d}_{t}) + 2\lambda \mathbf{d}_{i} \nonumber \\
&=& \Bigg(\frac{-e^{-(\mathbf{d}_{i}^{T}\mathbf{d}_{j} - \mathbf{d}_{i}^{T}\mathbf{d}_{t})}}{1 + e^{-(\mathbf{d}_{i}^{T}\mathbf{d}_{j} - \mathbf{d}_{i}^{T}\mathbf{d}_{t})}} \Bigg)(\mathbf{d}_{j} - \mathbf{d}_{t}) + 2\lambda \mathbf{d}_{i} \nonumber \\
\label{eq:7}
\end{eqnarray}

Using the similar chain rule derivation, the gradient of the objective function $OBJ$ w.r.t. $\mathbf{d}_{j}$ and $\mathbf{d}_{t}$ can be obtained as follows:

\begin{equation}
\frac{\partial OBJ}{\partial \mathbf{d}_{j}}  
= \Bigg(\frac{-e^{-(\mathbf{d}_{i}^{T}\mathbf{d}_{j} - \mathbf{d}_{i}^{T}\mathbf{d}_{t})}}{1 + e^{-(\mathbf{d}_{i}^{T}\mathbf{d}_{j} - \mathbf{d}_{i}^{T}\mathbf{d}_{t})}} \Bigg) \times \mathbf{d}_{i} + 2\lambda \mathbf{d}_{j} 
\label{eq:8}
\end{equation}

\begin{equation}
\frac{\partial OBJ}{\partial \mathbf{d}_{t}} 
= \Bigg(\frac{-e^{-(\mathbf{d}_{i}^{T}\mathbf{d}_{j} - \mathbf{d}_{i}^{T}\mathbf{d}_{t})}}{1 + e^{-(\mathbf{d}_{i}^{T}\mathbf{d}_{j} - \mathbf{d}_{i}^{T}\mathbf{d}_{t})}} \Bigg) \times (-\mathbf{d}_{i}) + 2\lambda \mathbf{d}_{t} 
\label{eq:9}
\end{equation}

Then embedding vectors $\mathbf{d}_{i}$, $\mathbf{d}_{j}$, and $\mathbf{d}_{t}$ are updated as below:

\begin{eqnarray}
\mathbf{d}_{i} = \mathbf{d}_{i} - \alpha \frac{\partial OBJ}{\partial \mathbf{d}_{i}} \nonumber \\
\mathbf{d}_{j} = \mathbf{d}_{j} - \alpha \frac{\partial OBJ}{\partial \mathbf{d}_{j}} \nonumber \\
\mathbf{d}_{t} = \mathbf{d}_{t} - \alpha \frac{\partial OBJ}{\partial \mathbf{d}_{t}} \nonumber \\
\label{eq:10}
\end{eqnarray}

where $\alpha$ is the learning rate.

Likewise, when the training instances come from person-person network, and person-document bipartite network, we update their corresponding gradients accordingly. We omit the detailed derivations here since they are very similar to the aforementioned ones.

\begin{algorithm}
\renewcommand{\algorithmicrequire}{\textbf{Input:}}
\renewcommand{\algorithmicensure}{\textbf{Output:}}
\caption{Network Embedding based Name Disambiguation in Anonymized Graphs}
\label{alg:1} 
\begin{algorithmic}[1]
\REQUIRE name reference $a$, dimension $k$, $\lambda$, $\alpha$, $L$
\ENSURE document embedding matrix $\mathbf{D}$ and its clustering membership set $\mathcal{C}$
\STATE Given name reference $a$, construct its associated $\mathcal{D}^{a}$, $\mathcal{A}^{a}$, $G_{pp}$, $G_{pd}$, $G_{dd}$
\STATE Given $G_{pp}$, $G_{pd}$, $G_{dd}$, construct training sample sets $\mathcal{P}_{G_{pp}}$, $\mathcal{N}_{G_{pp}}$, $\mathcal{P}_{G_{pd}}$, $\mathcal{N}_{G_{pd}}$, $\mathcal{P}_{G_{dd}}$, $\mathcal{N}_{G_{dd}}$ respectively based on edge sampling and uniform sampling techniques
\STATE Initialize $\mathbf{A}$ and $\mathbf{D}$ as $k$-dimensional matrices
\FOR {each training instance in training sample sets}
  \STATE Update involved parameters using SGD as described in Section~\ref{sec:modellearning}
\ENDFOR
\STATE Given $\mathbf{D}$ and $L$, perform HAC to partition $N$ documents in $\mathcal{D}^{a}$ into $L$ disjoint sets for name disambiguation
\STATE \textbf{return} $\mathbf{D}$, $\mathcal{C} = \{c_{1}, c_{2}, ..., c_{N} \}$
\end{algorithmic}  
\end{algorithm}

\subsection{Pseudo-code and Complexity Analysis }

The pseudo-code of the proposed network embedding method for name disambiguation 
under anonymized graphs is summarized in Algorithm~\ref{alg:1}.
The entire process consists of two phases: network embedding for document representation and name disambiguation
by clustering. Specifically, 
given a name reference $a$ and its associated document set $\mathcal{D}^{a}$ we aim to disambiguate, we first prepare the training instances in Line 1-2. 
Line 3 initializes the person and document embedding matrices $\mathbf{A}$ and $\mathbf{D}$ by randomly sampling elements from uniform distribution $[-0.2, 0.2]$. Then we train our proposed network embedding model
and update $\mathbf{A}$ and $\mathbf{D}$ using the training samples based on the SGD optimization in Line 4-6. Then given the obtained document embedding matrix $\mathbf{D}$ and $L$, in Line 7, we perform HAC to partition $N$ documents in $\mathcal{D}^{a}$ into $L$ disjoint sets such that each partition contains documents of a unique person entity with name reference $a$. Finally in Line 8, we return document embedding matrix $\mathbf{D}$ and its clustering membership set $\mathcal{C} = \{c_{1}, ..., c_{i}, ..., c_{N}\}$ for evaluation, where $1 \le c_{i} \le L$.

For the time complexity analysis, for the document embedding, when the training sample is $(d_{i}, d_{j}) \in \mathcal{P}_{G_{dd}}$, as  observed from Equations~\ref{eq:7}, ~\ref{eq:8} and~\ref{eq:10}, the cost
of calculating gradient of $OBJ$ w.r.t. $\mathbf{d}_{i}$ and $\mathbf{d}_{j}$, and updating $\mathbf{d}_{i}$ and $\mathbf{d}_{j}$ are both $\mathcal{O}(k)$. Similar analysis can be applied when
training instances are from $\mathcal{P}_{G_{pp}}$, $\mathcal{N}_{G_{pp}}$, $\mathcal{P}_{G_{pd}}$, $\mathcal{N}_{G_{pd}}$, $\mathcal{N}_{G_{dd}}$. Therefore, the total computational cost is 
$\big(2 * |\mathcal{P}_{G_{pp}}| + 2 * |\mathcal{P}_{G_{pd}}| + 2 * |\mathcal{P}_{G_{dd}}|\big)\mathcal{O}(k)$. For the name disambiguation, the computational cost of hierarchical clustering
is $\mathcal{O}(N^{2}logN)$~\cite{Zaki.Wagner:14}. So the total computational complexity of Algorithm~\ref{alg:1} is $\big(2 * |\mathcal{P}_{G_{pp}}| + 2 * |\mathcal{P}_{G_{pd}}| + 2 * |\mathcal{P}_{G_{dd}}|\big)\mathcal{O}(k) + \mathcal{O}(N^{2}logN)$.

%
%

\section{Experiments and Results}

We perform several experiments to validate the performance of our proposed network embedding method
for solving the name disambiguation task in a privacy-preserving setting using only linked data. 
We also compare our method with various other methods to demonstrate its superiority over those methods.

\subsection{Datasets}

A key challenge for the evaluation of name disambiguation task is the lack of availability of
labeled datasets from diverse application domains. In recent years, the bibliographic repository sites, Arnetminer~\footnote{\url{https://aminer.org/disambiguation}} and CiteSeerX~\footnote{\url{http://clgiles.ist.psu.edu/data/}} have published several ambiguous author name 
references along with respective ground truths (paper list of each real-life author), which we use for 
evaluation. From each of these two sources, we use $10$ highly ambiguous (having
a larger number of distinct authors for a given name) name references and show the performance of
our method on these name references. The statistics of name references in Arnetminer and CiteSeerX datasets 
are shown in Table~\ref{tab:arnet} and~ Table~\ref{tab:citeseerx}, respectively. In these tables, for each name reference, 
we show the number of documents, and the number of distinct authors associated with that name
reference. It is important to understand that the name disambiguation model is built on a name
reference, not on a source dataset such as, Arnetminer or CiteSeerX as a whole, so each name reference is a distinct dataset on which the evaluation is performed.

\begin{table}[t!]
\centering
\scalebox{0.90}{
\begin{tabular}{c c c}
\toprule
Name Reference & \# Documents & \# Distinct Authors \\ \midrule
Jing Zhang & 160 & 33 \\
Bin Yu & 78 & 8 \\
Rakesh Kumar & 82 & 5 \\
Lei Wang & 222 & 48 \\
Bin Li & 135 & 14 \\
Yang Wang & 134 & 23 \\
Bo Liu & 93 & 19 \\
Yu Zhang & 156 & 26 \\
David Brown & 42 & 9 \\
Wei Xu & 111 & 21 \\
\bottomrule
\end{tabular}}
\caption{Arnetminer Name Disambiguation Dataset}
\label{tab:arnet}
\vspace{-0.10in}
\end{table}

\begin{table}[t!]
\centering
\scalebox{0.90}{
\begin{tabular}{c c c}
\toprule
Name Reference & \# Documents & \# Distinct Authors \\ \midrule
K Tanaka & 174 & 9 \\
M Jones & 191  & 10 \\
J Smith & 798 & 26 \\
Y Chen & 848 & 64 \\
J Martin & 51 & 13 \\
A Kumar & 149 & 10 \\
J Robinson & 123 & 9 \\
M Brown & 118 & 13 \\
J Lee & 891 & 93 \\
S Lee & 1091 & 74 \\
\bottomrule
\end{tabular}}
\caption{CiteSeerX Name Disambiguation Dataset}
\label{tab:citeseerx}
\vspace{-0.20in}
\end{table}

%
%

\subsection{Competing Methods}

%
%

To validate the disambiguation performance of our proposed approach, we compare it against $9$ different methods. For a fair
comparison, all of these methods accommodate the name disambiguation using only relational data. Among all the competing methods, Rand, AuthorList, and AuthorList-NNMF are a set of primitive baselines that we have designed. But,
the remaining methods are taken from recently published works. For instance,
GF, DeepWalk, LINE, Node2Vec, and PTE are existing state-of-the-art approaches for vertex embedding, which we use for name disambiguation by clustering the documents using HAC in the embedding space similar to our approach.
Graphlet based graph kernel methods (GL3, GL4) are existing state-of-the-art approaches for name disambiguation in anonymized graphs. More details of each of the competing methods are given below. For each method, for a given name reference, a list of documents need to be partitioned among $L$ (user defined) different clusters.

\noindent \textbf{(1) Rand:} This naive method randomly assigns one of existing classes to the associated documents.\\
\textbf{(2) AuthorList:} Given the associated documents, we first aggregate the author-list of all documents in an 
author-array, then define a binary feature for each author, indicating his presence or absence in the author-list of that document. Finally we use HAC with the generated author-list as features for disambiguation task. \\ 
\textbf{(3) AuthorList-NNMF:} We perform Non-Negative Matrix Factorization (NNMF) on the generated author-list features the same way described above. Then the latent features from NNMF are used in a HAC framework for disambiguation task.\\
\textbf{(4) Graph Factorization (GF)~\cite{Kuang.Ding.ea:12}:} We first represent co-authorship network $G_{pp}$ and the linked document network $G_{dd}$ as affinity matrices, and then utilize matrix factorization technique to represent each document into low-dimensional vector. Note that GF is optimized via a point-wise regression model that minimizes a square loss function. However, in our proposed embedding approach, the objective aims to minimize a ranking loss function, which is substantially different from GF. \\
\textbf{(5) DeepWalk~\cite{Perozzi.Skiena.ea:14}:} DeepWalk is an approach recently proposed for network embedding, which is only applicable for homogeneous network with binary edges. Given $G_{pp}$ and $G_{dd}$, we use uniform random walk to obtain the contextual information of its neighborhood for document embedding~\footnote{Code is available at \url{http://www.perozzi.net/projects/deepwalk/}}.\\ 
\textbf{(6) LINE~\cite{Tang.Wang.ea:15}:} LINE aims to learn the document embedding that preserves both the first-order 
and second-order proximities~\footnote{Implementation Code is available at \url{https://github.com/tangjianpku/LINE}}. Note that LINE can only handle the embedding of homogeneous network and the embedding formulation and optimization are quite different from the one proposed in our work. \\ 
\textbf{(7) Node2Vec~\cite{Grover.Leskovec:16}:} Similar to DeepWalk, Node2Vec designs a biased random walk procedure 
for document embedding.~\footnote{We use the code from \url{https://github.com/aditya-grover/node2vec}}.\\ 
\textbf{(8) PTE~\cite{Tang.Qu.ea:15}:} Predictive Text Embedding (PTE) framework aims to capture the relations of word-word, word-document, and word-label. However, such keyword and label based biographical features are not available in the anonymized setup. Instead we utilize local structural information of both $G_{pp}$ and $G_{pd}$ networks to learn the document embedding. However, this approach is not able to capture the linked information among documents.\\  
\textbf{(9) Graph Kernel~\cite{Hermansson.Kerola.ea:13}:} In this work, size-3 graphlets (GL3) and size-4 graphlets (GL4) 
are used to build graph kernels, which measure the similarity between documents. Then the learned similarity metric is 
used as features in HAC for name disambiguation. As we see, both kernels only use network topological information.~\footnote{The kernel values are obtained by source code supplied by the original authors}

\subsection{Experimental Setting and Implementation}

For each of the $20$ name references, we perform name disambiguation task using our proposed
method and each of the competing methods to demonstrate that our proposed method is superior
than the competing methods. For evaluation metric, we use Macro-F1 measure~\cite{Zaki.Wagner:14}, 
which is the unweighted average of F1 measure of each class. 
The range of Macro-F1 measure is between $0$ and $1$, and a higher value indicates better 
disambiguation performance. Besides comparison with competing methodologies, we also perform
experiments to show that our method is robust against the variation of user defined parameters
(specifically, embedding dimension and the number of clusters) over a wide range of parameter 
values. Experiments are also performed to show how the embedding model performs with each of the three types of networks (person-person, person-document, and document-document) incrementally added. Finally, we show the convergence of
the learning model while performing the document embedding phase. 

There are a few user defined parameters in our proposed embedding model. The first among these is the embedding dimension $k$, which we set to be $20$. For the regularization parameter in model inference (see Section~\ref{sec:modellearning}), we perform grid search on the validation set in the following range: $\lambda = \{0.001, 0.005, 0.01, 0.1, 1, 10\}$. In addition to that, we fix the learning rate $\alpha = 0.02$. For the disambiguation stage, we use the actual number of classes $L$ of each name reference as input to perform HAC.
For both data processing and model implementation, we implement our own code in Python and use
NumPy, SciPy, scikit-learn, and Networkx libraries for linear algebra, machine learning, and graph operations. 
We run all the experiments on a 2.1 GHz Machine with 8GB memory running Linux operating system.

\begin{table*}[t!]
\centering
\scalebox{0.85}{
\begin{tabular}{c | c | c  c  c |  c  c  c  c  c | c c | c}
\toprule
Name  &  \textbf{Our Method} &  Rand & AuthorList & AuthorList- & GF~\cite{Kuang.Ding.ea:12} & DeepWalk~\cite{Perozzi.Skiena.ea:14} & LINE~\cite{Tang.Wang.ea:15} & Node2Vec~\cite{Grover.Leskovec:16} & PTE~\cite{Tang.Qu.ea:15} & GL3~\cite{Hermansson.Kerola.ea:13}  & GL4~\cite{Hermansson.Kerola.ea:13} & Improv.\\
Reference & & &  & NNMF & & &  & &  &  \\
\midrule
Jing Zhang & \textbf{0.734 (0.014)} & 0.192  & 0.327 & 0.463 & 0.669 & 0.654 & 0.651 & 0.312 & 0.458 & 0.318 &  0.329 & 9.7\% \\
Bin Yu & \textbf{0.804 (0.009)} & 0.201 & 0.371 & 0.283  & 0.610 & 0.644 & 0.643  & 0.531 & 0.399 & 0.489  & 0.504 & 24.8\% \\
Rakesh Kumar & \textbf{0.834 (0.012)} & 0.226 & 0.305 & 0.404 & 0.448 & 0.617 & 0.641 & 0.372 & 0.219 & 0.434 & 0.407 & 30.1\% \\
Lei Wang & \textbf{0.805 (0.021)} & 0.198 & 0.502 & 0.424  & 0.633 & 0.419 & 0.639 & 0.263 & 0.447 & 0.291  & 0.321 & 26.0\% \\
Bin Li & \textbf{0.848 (0.016)} & 0.172 & 0.610 & 0.733 & 0.761 & 0.392 & 0.641 & 0.186 & 0.349 & 0.336  & 0.418 & 11.4\% \\
Yang Wang & \textbf{0.798 (0.011)} &  0.199 & 0.442 & 0.532 & 0.575 & 0.640 & 0.623 & 0.331 & 0.444 &  0.378 &  0.512 & 24.7\% \\
Bo Liu & 0.831 (0.022) &  0.215 & 0.482 & 0.740 & \textbf{0.850} & 0.788 & 0.781 & 0.459 & 0.373 &  0.498 & 0.347 & -2.2\%  \\
Yu Zhang & \textbf{0.820 (0.031)} & 0.186 & 0.519 & 0.566 & 0.565 & 0.454 & 0.658 & 0.196 & 0.385 & 0.369 & 0.305 & 24.6\% \\
David Brown & \textbf{1.00 (0.00)} &  0.304 & 0.818 & 0.583 & 0.802 & 0.494 & \textbf{1.00} & 0.221 & 0.575 &  0.603 &  0.698 & 0\%\\
Wei Xu & \textbf{0.793 (0.014)}  & 0.256 & 0.527 & 0.564 & 0.625 & 0.228 & 0.599 & 0.136 & 0.236 &  0.386  & 0.428 & 26.9\% \\
\bottomrule
\end{tabular}}
\caption{Comparison of Macro-F1 values between our proposed method and other competing methods for name disambiguation task in Arnetminer dataset (embedding dimension = 20). Paired $t$-test is conducted on all performance comparisons and it shows that all improvements are significant at the $0.05$ level.}
\label{tab:result1}
\vspace{-0.10in}
\end{table*}

\begin{table*}[t!]
\centering
\scalebox{0.85}{
\begin{tabular}{c | c | c  c  c | c  c  c  c  c | c  c | c}
\toprule
Name  & \textbf{Our Method} & Rand & AuthorList & AuthorList- & GF~\cite{Kuang.Ding.ea:12} & DeepWalk~\cite{Perozzi.Skiena.ea:14} & LINE~\cite{Tang.Wang.ea:15} & Node2Vec~\cite{Grover.Leskovec:16} & PTE~\cite{Tang.Qu.ea:15} & GL3~\cite{Hermansson.Kerola.ea:13} & GL4~\cite{Hermansson.Kerola.ea:13} & Improv. \\
Reference &  &  &   &  NNMF & & &  & &  &  \\
\midrule
K Tanaka & \textbf{0.706 (0.018)} & 0.178  & 0.202 & 0.168 & 0.334 & 0.450 & 0.398 & 0.304 & 0.173 & 0.235  & 0.276 & 56.9\% \\
M Jones & \textbf{0.743 (0.009)} &  0.184 & 0.189 & 0.261 & 0.529 & 0.696 & 0.688 & 0.513 & 0.348 & 0.216  &  0.398 & 6.8\% \\
J Smith & \textbf{0.503 (0.007)} &  0.083 & 0.121 & 0.280 & 0.316 & 0.098 & 0.104 & 0.073 & 0.136 &  0.201 &  0.237 & 59.2\% \\
Y Chen & 0.367 (0.019) &  0.069 & 0.325 & 0.355 & \textbf{0.439} & 0.118 & 0.193 & 0.058 & 0.199 & 0.334 &  0.385 & -16.4\% \\
J Martin & \textbf{0.898 (0.021)}  &  0.310 & 0.624 & 0.536 & 0.755 & 0.728 & 0.774 & 0.629 & 0.587  & 0.414 & 0.431 & 16.0\% \\
A Kumar & \textbf{0.645 (0.006)} &  0.166 & 0.251 & 0.375 & 0.319 & 0.407 & 0.395 & 0.424 & 0.247 & 0.192  & 0.234 & 52.1\% \\
J Robinson & \textbf{0.796 (0.033)} &  0.200 & 0.348 & 0.438 & 0.393 & 0.513 & 0.603 & 0.608 & 0.345 & 0.271 & 0.316 & 30.9\% \\
M Brown & \textbf{0.741 (0.028)} &  0.171 & 0.306 & 0.573 & 0.478 & 0.481 & 0.633 & 0.211 & 0.269 & 0.297 & 0.248 & 17.1\%  \\
J Lee & 0.366 (0.038) &  0.089 & 0.262 & 0.256 & 0.231  & \textbf{0.387} & 0.134 & 0.181 & 0.142 &  0.189 &  0.205 & -5.4\% \\
S Lee & \textbf{0.624 (0.015)} &  0.057 & 0.214 & 0.248 & 0.345 & 0.194 & 0.109 & 0.044 & 0.074 &  0.215 &  0.268 & 80.9\% \\
\bottomrule
\end{tabular}}
\caption{Comparison of Macro-F1 values between our proposed method and other competing methods for name disambiguation task in CiteSeerX dataset (embedding dimension = 20). Paired $t$-test is conducted on all performance comparisons and it shows that all improvements are significant at the $0.05$ level.}
\label{tab:result2}
\vspace{-0.10in}
\end{table*}

\subsection{Comparison among Various Name Disambiguation Methods}

Table~\ref{tab:result1} and Table~\ref{tab:result2} show the performance comparison of
name disambiguation between our proposed method and other competing methods
for all 20 name references (one table for ArnetMiner names, and the other for CiteSeerX names). 
In both tables, the rows correspond to the name references and the columns ($2$ to $12$) stand
for various methods. The competing methods are grouped logically. The first group includes the 
baseline methods that we have designed such as random
predictor (Rand) and methods using low-dimensional factorization of author-list for clustering.
The second group includes various state-of-the-art network embedding methodologies, and the third
group includes two methods using graphlet based graph kernels. The cell values are the performance 
of a method using Macro-F1 score for disambiguation of documents under a given name reference.
The last column shows the overall improvement of our proposed method compared with the best competing method. 
Since SGD based optimization technique in our proposed 
embedding model is a randomized method, for each name reference we execute the method $10$ times 
and report the average Macro-F1 score. For our method, we also show the standard deviation in the parenthesis.~\footnote{Standard deviation for other competing methods are not shown due to the 
space limit.} For better visual comparison, we highlight the best Macro-F1 score of each name reference 
with bold-face font. 

As we observe,  our proposed embedding model performs the best
for $9$ and $8$ name references (out of $10$) in Table~\ref{tab:result1}, and Table~\ref{tab:result2}, respectively.
Besides, the overall percentage
improvement that our method delivers over the second best method is relatively large. 
For an example, consider the name
``S Lee" shown in the last row of Table~\ref{tab:result2}. This is a difficult disambiguation task; from  Table~\ref{tab:citeseerx}, it has 1091 documents and 74 distinct real-life authors ! A random
predictor (Rand) obtains a Macro-F1 of only 0.057 due to the large number of classes. Whereas our
method achieves $0.624$ Macro-F1 score for this name reference; the second best method for this name (GF) achieves 
only $0.345$, indicating a substantial improvement ($80.9\%$) by our method. The 
relatively good performance of our
proposed method across all the name references is due to the fact that the method is able to learn document embedding,
which is particularly suited for the name disambiguation task 
 by facilitating information exchange among the three networks (see Section~\ref{sec:pf}).

Among the competing methods, AuthorList based methods perform poorly because
the binary features are not intelligent enough to disambiguate documents, even
after using traditional low dimensional embedding by non-negative matrix
factorization. Graph kernel based methods such as GL3 and GL4 also have similar
fate; the possible reason could be that the size-3 and size-4 graphlet
structures are not decisive patterns to distinguish documents authored by
different persons. On the other hand, embedding based methods are much better
as they are able to learn effective features, which bring the documents
authored by the same real-life person in close proximity in the feature space.
This finding justifies our approach of choosing a document embedding method for
solving name disambiguation. Among the competing network embedding based
approaches, as we can observe from all name references, no single method
emerges as a clear winner. To be more precise, PTE performs poorly as it fails
to incorporate linked structural information among the documents. Both GF and
LINE outperform DeepWalk in majority of name references. This is because
DeepWalk ignores the weights of the edges, which is considered to be very
important in the linked document network. However, neither of embedding based
competing methods could encode the document co-occurrence by exploiting the
information from multiple networks, which is exploited by our proposed model. 
Besides, as mentioned earlier, our similarity ranking based objective function is 
better suited than the K-L divergence based objective functions for placing the 
nodes in the embedding space for facilitating a downstream clustering task. This
is possibly a significant reason
for our method to show superior performance
over the existing network embedding based methods.

%
%

\subsection{Parameter Sensitivity of Embedding Dimension}

\begin{figure}[t]
\centering
\includegraphics[width=60mm]{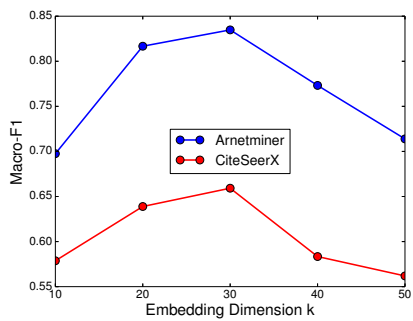}
\caption{The effects of embedding dimension on the name disambiguation performance}
\label{fig:latent}
\end{figure}

We also perform experiment to show how the embedding dimension $k$ affects the
disambiguation performance of our proposed method. Specifically, we vary the
number of embedding dimension $k$ as $\{10, 20, 30, 40, 50 \}$. For the sake of
space, in each of the datasets, we show the average results over all the $10$
name references. The disambiguation results are given in
Figure~\ref{fig:latent}. As we observe, for both datasets, as the dimension of
embeddings increases, the disambiguation performance in terms of Macro-F1 first
increases and then decreases. The possible explanation could be that
when the embedding dimension is too small, the embedding representation capability is not sufficient. However, when
the embedding dimension is too large, the proposed embedding model may overfit the data, leading to the unsatisfactory
disambiguation performance.

\subsection{Performance Comparison over the Number of Clusters}

\begin{figure}[t]
\centering
\includegraphics[width=60mm]{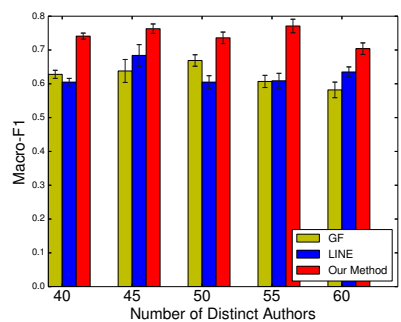}
\caption{Macro-F1 results of multiple $L$ values on name reference ``Lei Wang" using Our Method, GF, and LINE (embedding dimension = 20).}
\label{fig:casestudy}
\end{figure}

One of the potential problems for name disambiguation is to determine the number of real-life persons $L$ under a given name reference, because in real-life $L$ is generally unknown a-priori. So a method whose performance is superior
over a range of $L$ values should be preferred. For this comparison, after learning the document representation, we use various $L$ values as input in the HAC for name disambiguation and record the Macro-F1 score over different $L$
for the competing methods. In our experiment, we compare Macro-F1 value of our method with two other best performing methods over several names, but due to space limitation, we show this result only for one 
name (``Lei Wang'' in Arnetminer) using bar-charts in Figure~\ref{fig:casestudy}. In this figure, we compare the performance differences between our method with two other best performing methods (GF and LINE) as we vary $L$ as $\{40, 45, 50, 55 ,60 \}$. Note that the actual number of distinct authors under ``Lei Wang" is $48$ as shown in Table~\ref{tab:arnet}. As we can see, our proposed method always outperforms the state-of-the-art with all different $L$ values, and the overall improvement of our method over these two methods is statistically significant with a $p$-value of less than $0.01$. Because of the robustness of our proposed embedding method for name disambiguation regardless of $L$ values,
this is a better method for the real-life application. 

%
%

\subsection{Component Contribution Analysis}

\begin{figure}[t]
\centering
\includegraphics[width=60mm]{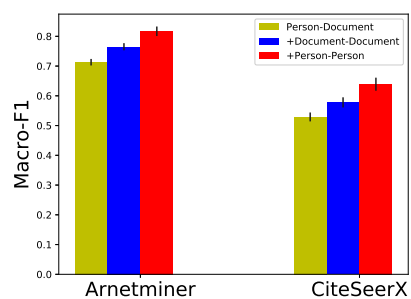}
\caption{Component Contribution Analysis in terms of name disambiguation performance using Arnetminer and CiteSeerX as a whole source (embedding dimension = $20$).}
\label{fig:comp_analysis}
\end{figure}

Our proposed network embedding model is composed of three types of
networks, namely person-person, person-document, and linked document networks
(explained in Section~\ref{sec:pf}). In this section we study the contribution
of each of the three components for the task of name disambiguation by incrementally
adding the components in the network embedding model.  Specifically,
we first rank each individual component by its disambiguation performance in
terms of Macro-F1, then add the components one by one in the order of their
disambiguation power. In particular, we first add person-document graph,
followed by linked document graph, and person-person graph.
Figure~\ref{fig:comp_analysis} shows the name disambiguation performance in
terms of Macro-F1 value using our proposed network embedding model with
different component combinations. As we see from the figure, after adding each
component, we observe improvements for both datasets, in which the results are
averaged out over all the $10$ name references.

\subsection{Convergence Analysis}

\begin{figure}[!t]
\centering
\subfigure[Loss vs Epoch]{\label{fig:a}\includegraphics[width=40mm]{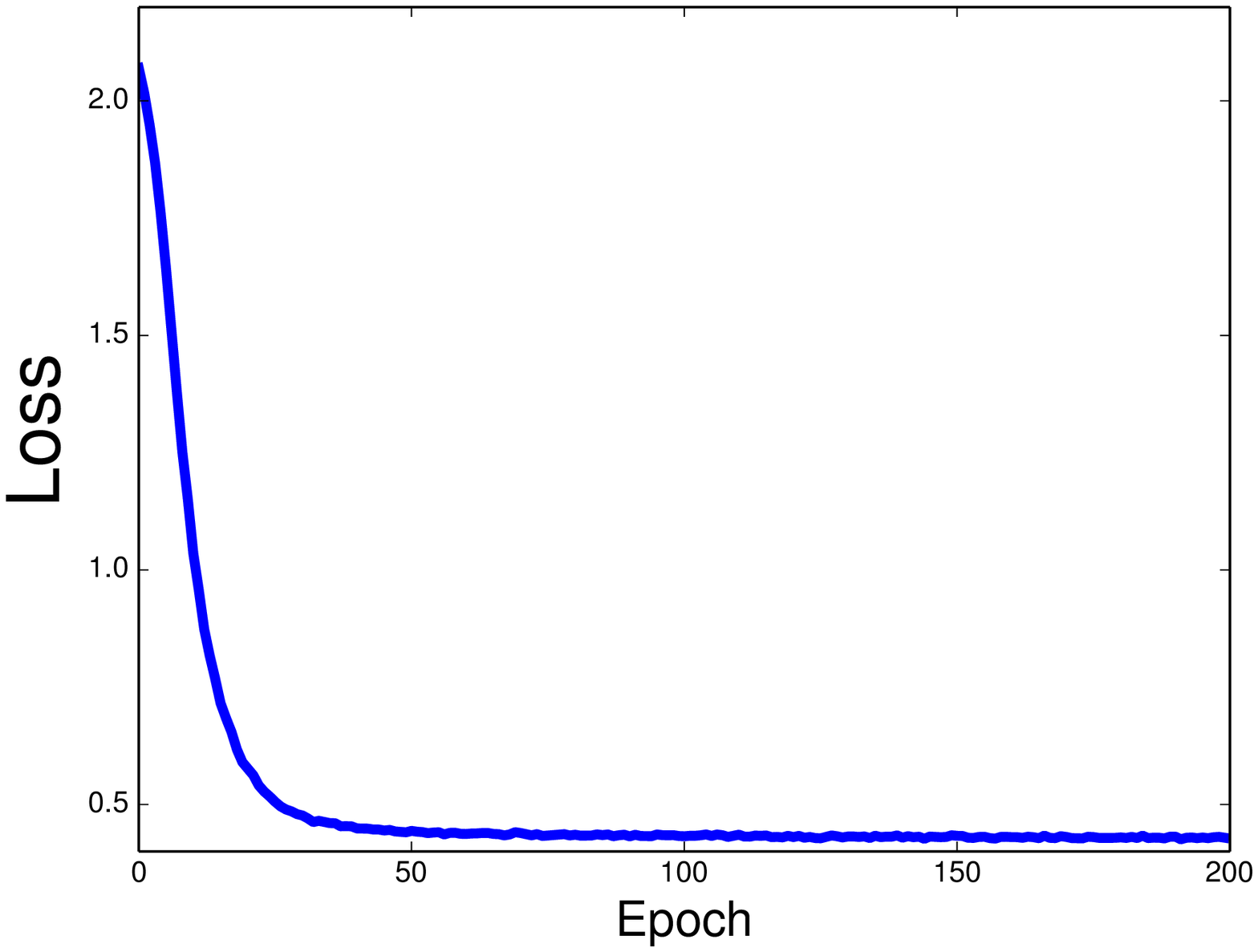}}
\subfigure[AUC vs Epoch]{\label{fig:d}\includegraphics[width=40mm]{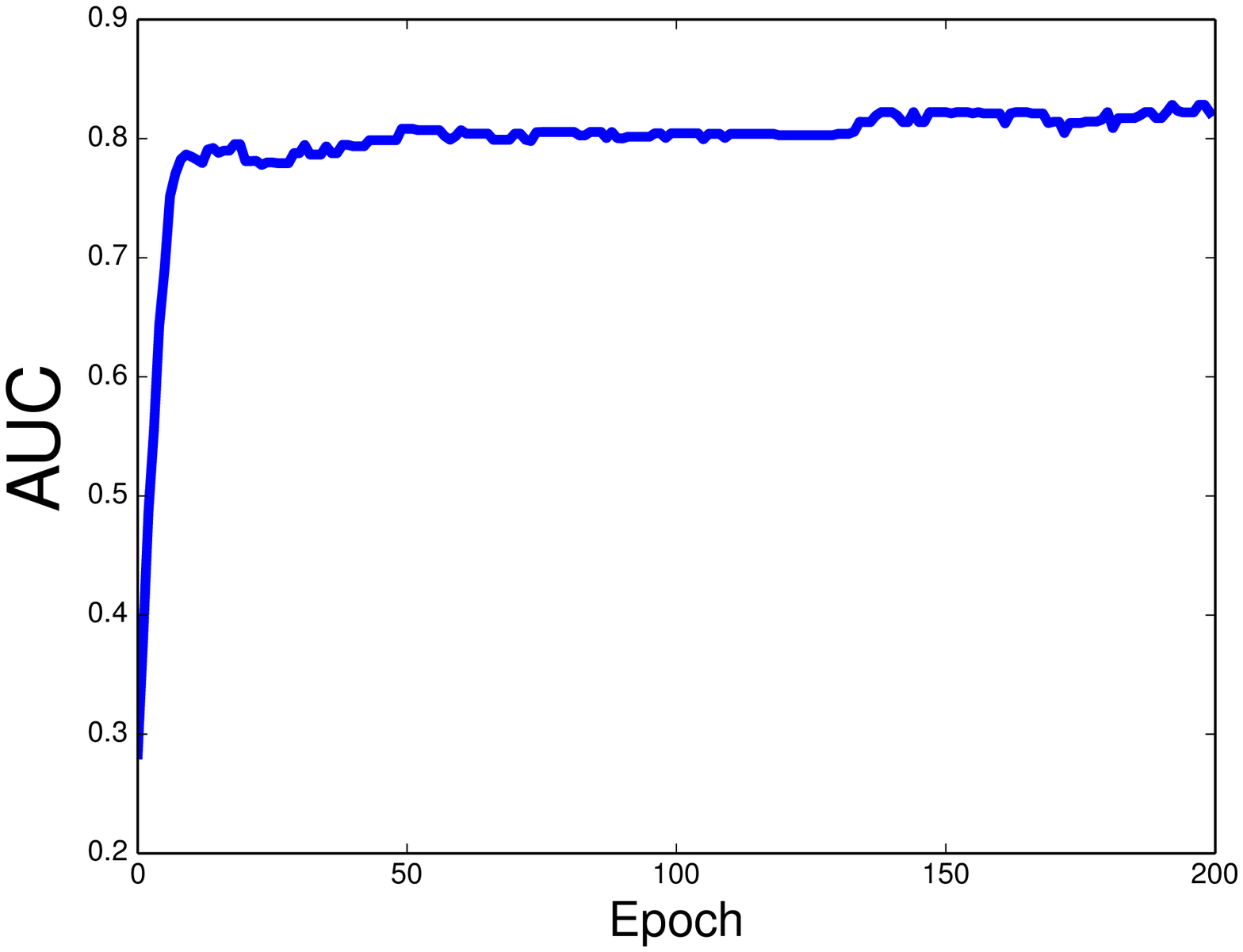}}
\caption{Convergence analysis in terms of both objective loss and AUC of name reference ``Lei Wang'' using our proposed network embedding model for name disambiguation.}\label{fig:convergence}
\label{fig:conver}
\end{figure}

We further investigate the convergence of proposed network embedding algorithm shown in Section~\ref{sec:method}. Figure~\ref{fig:conver} shows
the convergence analysis of our method under the name reference ``Lei Wang" from Arnetminer. For each epoch, we sample $\bigg(\left|E_{pp}\right| + \left|E_{pd}\right| + \left|E_{dd}\right|\bigg)$ training instances to update the corresponding model embedding vectors. We can observe that our proposed network embedding approach converges approximately within
$50$ epochs and achieves promising convergence results on both pairwise ranking based objective loss and AUC. However, as shown in Equation~\ref{eq:6}, the objective function in our proposed embedding model is not convex, thus reaching global optimal solution using SGD based optimization technique is a fairly challenging task. The possible remedy could be to decrease the learning rate $\alpha$ in SGD when number of epochs increases. Another strategy is to try multiple runs with different seeds initialization. Similar convergence patterns are observed for other name references as well. 

\section{Conclusion}

To conclude, in this paper we propose a novel representation learning based
solution to address the name disambiguation problem. Our proposed representation
learning model uses a pairwise ranking objective function which clusters the
documents belonging to a single person better than other existing network
embedding methods.  Besides, the proposed solution uses only the relational
data, so it is particularly useful for name disambiguation in anonymized
network, where node attributes are not available due to the privacy concern.
Our experimental results on multiple datasets show that our proposed method
significantly outperforms many of the existing state-of-the-arts for name
disambiguation.

\bibliographystyle{ACM-Reference-Format}
\balance
\bibliography{disambiguation}  

\end{document}